# Fabrication of optically smooth Sn thin films

Vamsi Borra[1], Daniel G. Georgiev[1] and Corey R. Grice[2]

[1] Department of EECS, University of Toledo, Toledo, OH 43606-3390, U.S.A.

[2] Department of Physics and Astronomy, University of Toledo, Toledo, OH 43606-3390, U.S.A

Keywords: *Sn films, optical smoothness, thermal evaporation, SEM, AFM*

**Abstract**

The fabrication of optically smooth thin Sn films by vacuum or electrodeposition techniques is usually challenging. Little has been published on how to address this challenge mainly because very few applications require such smooth Sn surfaces. The excitation of surface plasmon polaritons on Sn surfaces by prism-based methods represents a case that requires very smooth surfaces and has motivated this work. It is shown that the deposition rate and the substrate temperature of a vacuum evaporation method can be optimized to obtain very smooth Sn films and this is supported by direct imaging evidence from atomic force microscopy and scanning electron microscopy.

## 1. Introduction

Optical measurements and experiments typically require high-quality smooth metal surfaces. Of particular interest to this work is the possibility of exciting surface plasmon polaritons (SPPs) on Sn films by near-field optical methods [1][2][3]. This requires optically smooth, chemically pure and stable Sn films, the fabrication of which turns out to be challenging if standard deposition techniques, such as vacuum based ones, sputtering or electroplating, are to be used. The main reason for this appears to be the combination of high chemical reactivity, low melting point, and the existence of different metastable crystalline phases, which characterizes Sn as well as other metals such as Zn or Cd.

There is an overwhelming amount of published work and thorough understanding on how to fabricate high-quality optically smooth films of more technologically important metals such as Au, Ag, Cu , but we have found very little on the fabrication of optically smooth thin Sn films using vacuum based methods. A review article by Muggleton [4] on the preparation of nuclear targets describes a method of deposition which involves *evaporation*, assisted by release agents. In this method, a releasing agent such as $BaCl_2$ or Teepol is coated on pyrex glass for self-supporting targets. However, there is no mention of the surface roughness in this work [5]. Sn evaporated from a molybdenum boat usually result in opaque films that have cloudy surfaces of low reflectivity as reported in [6]. Those films generally become coarser as the film thickness increases. MacRae *et al*. [7] used explosive evaporation of Sn onto quartz substrates (deposition rates of about 100 nm/sec) to study the optical properties of white tin (β phase), the obtained film surfaces turned out to be of poor quality despite being only 250nm thick. In a report by Golovashkin *et al*.

[8] relatively thick films were deposited on polished glass using vacuum evaporation. However, the evaporation conditions that were used were never fully disclosed, and the exact thickness obtained was not mentioned. It was also reported that annealing the samples for several hours in vacuum after the deposition had no effect on the films' quality.

Sn thin films obtained by other deposition methods such as *electrodeposition* and *sputtering* are reported by several groups. The electrodeposited Sn thin films reported by both [9] and [10] resulted in very rough surfaces. Sn thin films deposited by a sputtering method [11,12], also turned out to have significant roughness - an order of magnitude higher than that of evaporated films.

In this work, we re-examine the possibility of obtaining optically smooth Sn films by vacuum evaporation while relying on direct evidence from imaging techniques and we provide details and steps to be followed to obtain smooth Sn films.

## 2. Experimental

*2.1 Substrate preparation*

Both Si wafers and microscope soda-lime glass slides were used as substrate materials. Although the former type is smoother than the later one, we were able to show that it is possible to get optical smoothness on the relatively uneven microscope glass slides. Since Si wafers come pre-cleaned when compared to microscope slides, the cleaning needed is relatively little. However, we conducted several different cleaning trials and explored different cleaning combinations for the glass slides in order, for example, to eliminate or reduce the commonly observed streaks and other residue that are sometimes observed after performing evaporation.

Our general substrate cleaning procedure is as follows: wash in cleaning solution (Micro-90), then rinse with DI water, followed by ultra-sonication bath in methanol for 20-25 min, and, finally, ultra-sonication bath in ethanol for 20-25 min. In between these steps, the surfaces are rubbed with lint-free wipe and blown dry with nitrogen.

*2.2 Evaporation and substrate setup*

Sn thin films were prepared by vacuum evaporation of 99.999% pure Sn (from Kurt J. Lesker) from tungsten boats onto the Si and microscope slides simultaneously. The evaporation was carried out in a Denton Vacuum machine (model DV-502A) at base vacuum of $(3.5 – 4.8) \times 10^{-6}$ Pa. During the evaporation, the pressure raises to the $10^{-5}$ Pa range. Details about the evaporation process, the rate of deposition and substrate temperature needed for obtaining optically smooth films, are discussed in more detail next. Before the actual evaporation, a vacuum chamber cleaning procedure is performed by gradually increasing the current through an empty tungsten boat to a high value of 250 A and keeping it on for at least 10 min. Followed by a cooling step, the evaporator chamber is brought to atmospheric pressure by purging it with dry nitrogen. The 5N pure Sn pellet is then placed in the dimple tungsten boat along with the glass substrate for the actual evaporation. Once the pressure in the vacuum chamber reaches the $10^{-6}$ Pa range, the source shutter is kept closed while the current is gradually increased until the evaporation point is reached. The metal is let to boil for 30 – 45 sec before opening the shutter.

After performing several deposition runs and characterizing the surfaces of the obtained films, we observed that the surface roughness of the Sn films is strongly dependent on the rate of deposition. Deposition rates that are reported in literature [6], resulted in very rough films that have cloudy surfaces. Several deposition rate trials were performed on Si and microscope slide substrates to optimize the rates that are functional. Our optimized rates turned out to be between 2.6 – 2.8 nm/sec. More surface details are discussed in the results and discussion section.

We also observed that the temperature of the substrate is as important to obtaining smooth Sn films as is the optimized deposition rate. Substrates, placed on surfaces that are maintained at room temperature, or lower, during the evaporation turned out to be substantially smoother than the substrates that are on the hotter surfaces. The evaporation chamber that we used contains a glass slide rack, the temperature of which stays close to 21 ºC with no evaporation in progress and rises to only 38–43 ºC during evaporation. However, the temperature of the actual substrate holders rises to 100–107 ºC during the deposition. The reason that the glass rack does not change its temperatures significantly is because it is attached to a large metal block which is part of the evaporator's external body (maintained at room temperature). So, in our case we took advantage of the glass slide rack inside the evaporator chamber by placing our glass substrate on the rack instead of on the actual substrate holder. The distance between the Sn source in the tungsten boat and the substrates was about 20 cm.

*2.3 Film Characterization Methods*

The crystallographic properties of the prepared films were evaluated by grazing incidence X-ray diffraction (GIXRD) in a Rigaku Ultima III machine, using a Cu Kα radiation source. The surface morphology of the thin film samples was examined using a Nanoscope V atomic force microscope (AFM), operated in tapping-mode with a Si probe tip (OTESPA-R3, Bruker) and by scanning electron microscopy (SEM) in a Hitachi S-4800 machine, operated in secondary-electron mode with acceleration voltage of 5 kV in order to limit the observation only to the film. Compositional analysis was performed using an Oxford Instruments energy dispersive X-ray spectrograph (EDS), installed in the SEM system, with acceleration voltage of 20 kV. The EDS software (INCA) was calibrated using cobalt EDS standards prior to the EDS point & I.D. mapping. The thickness of the films was determined by step height measurements using a stylus profilometer (Veeco Dektak 150).

## 3. Results and discussion

Fig. 1 shows the XRD patterns obtained by GIXRD at 1º angle of incidence, taken from smooth (see also Fig. 2d) and rough (see Fig.2a) films on a glass substrate. The results suggest that the smooth films have a different degree of preferred orientation when compared to the rough films.

# Fabrication of optically smooth Sn thin films

For the smooth films, there is a higher degree of preferred orientation of the a-axis (of the β-Sn phase) in a direction perpendicular to the substrate surface as indicated by the larger peak-intensity ratio $I_{(200)}/I_{(101)}$ of 91% for the smooth films vs. 60.8% for the rough films. Also, there is a difference in the relative strength of the (301) reflection. Preferential growth was also observed by Hishita *et al.* [13] and Takeuchi *et al.*[14]. A possible explanation involving minimization of the surface energy by the (100) plane growth was suggested by Park *et al.* [15].

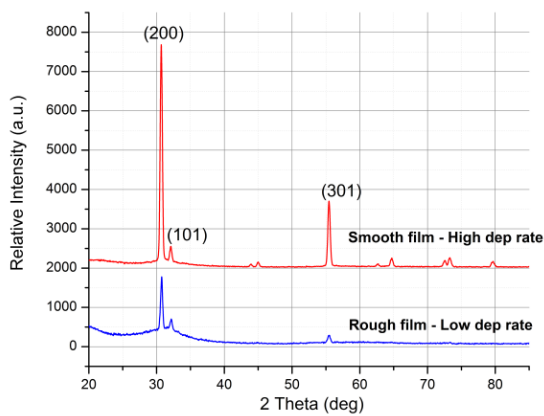

**Figure 1. XRD patterns for vacuum-evaporated Sn films on glass substrates. The smooth films are deposited at a rate of 2.6-2.8 nm/sec (in red color), and the rough films are deposited at 0.05-0.08 nm/sec (in blue).**

The films were examined by SEM to study the morphology and surface roughness; SEM images of the deposited films are shown in Fig. 2. The thickness of the deposited films was determined to be around 500 nm by using profilometry. The films that were deposited at 0.05 – 0.08 nm/s had very rough surfaces. An SEM scan of one such film is shown in Fig. 2(a). To the contrary, the films that were deposited by optimizing the deposition rate (2.6-2.8 nm/s) and maintaining the temperature of the substrate, resulted in a very smooth (optically smooth) surface as shown in the Fig. 2(d). Films obtained with deposition rates greater than 2.8 nm/s lead to very rough surfaces. The reason for this roughness at higher deposition rates could be related to the temperature raise of the substrates and the vacuum chamber. This temperature increase is caused by passing more current through the tungsten boat to attain higher deposition rates, which results in additional heating of the chamber. The optimal conditions were obtained after trying different deposition rates and conditions. SEM images of the surfaces of four films, representing four of these deposition condition combinations, are shown in Fig.2.

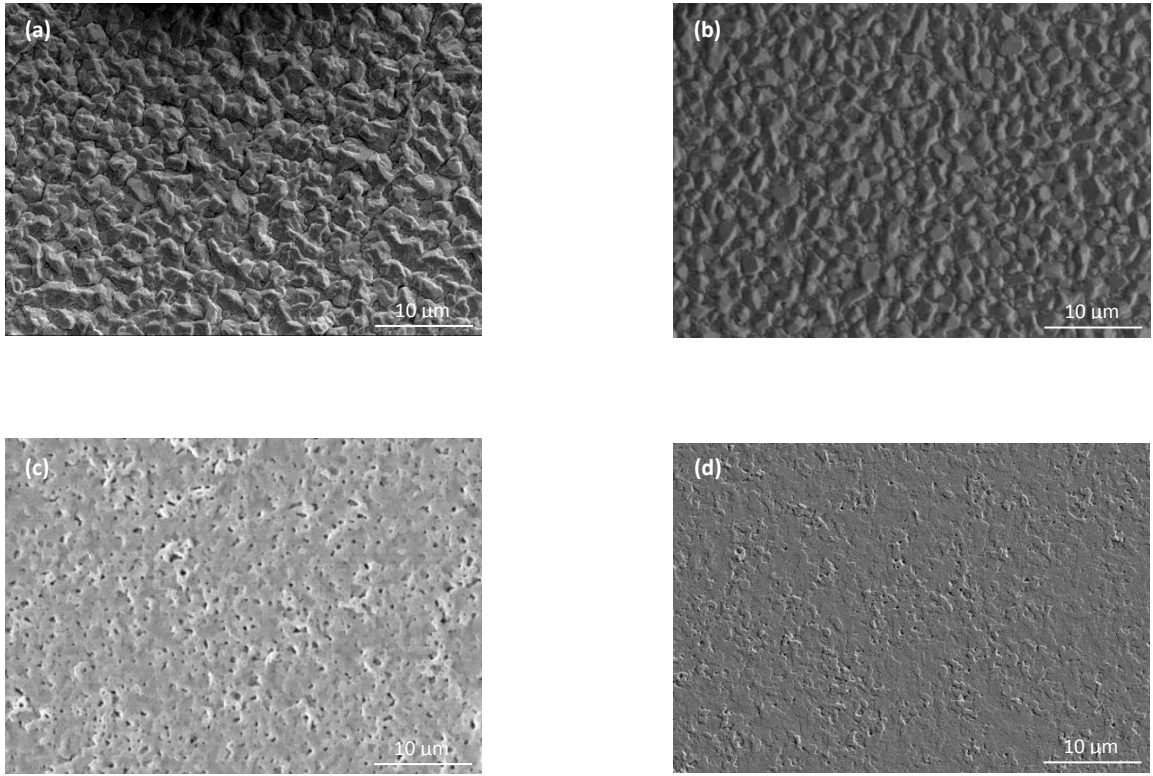

**Figure 2. SEM scans of the deposited Sn films on glass substrate (a) Deposited at 0.05-0.08 nm/s (b) Deposited at 0.08-0.15 nm/s (c) Deposited at 1.4-1.6 nm/s (d) Deposited at our optimized rates of 2.6-2.8 nm/s.**

A close-up view of the optically smooth film (Fig. 2(d)) is shown in Fig.3. This figure shows a clear sign of very flat Sn grains and its boundaries on the surface of the film. This smoothness can be attributed to the lack of time for grains to grow under these relatively high deposition rate conditions. The slower deposition, on the other hand, allows the grain growth leading to rough films (Fig. 2(a)).

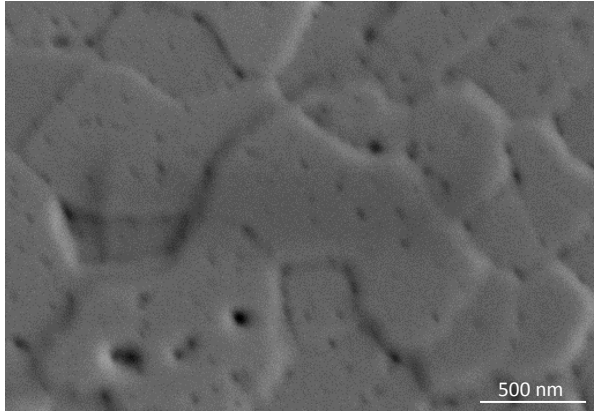

**Figure 3. Close-up SEM surface scan of the same (image Fig. 2(d)) smooth film (deposited at optimized rates of 2.6-2.8 nm/ s).**

The magnitude of the difference in the roughness of the films is somewhat less well seen in the larger area SEM scans shown in Fig. 4. AFM is used to complement the SEM work and to assess the roughness of the vacuum-evaporated films.

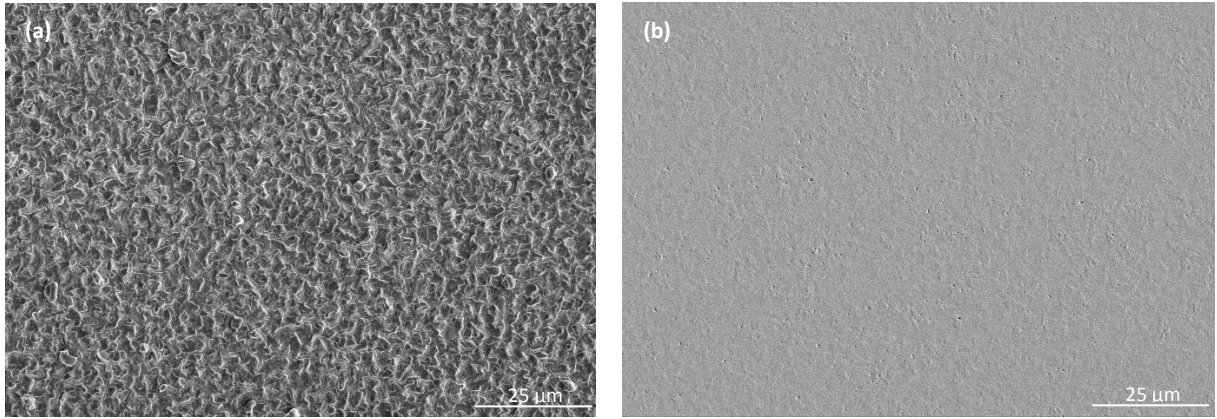

**Figure 4. Larger area SEM scans of the deposited films on glass substrate, (a) rough film (Deposited at 0.05-0.08 nm/s) and (b) smooth film (Deposited at optimized rates of 2.6-2.8 nm/s).**

Relatively large area (5 X 5 µm) AFM scans, taken from the smooth film (Fig. 2(d)) are shown in Fig. 5(a). The singular structures, such as narrow / broad cones or deep valleys, cannot be found on this large sample surface. Interestingly, the grain boundaries that are imaged using SEM (Fig. 3) are very clear in this 3D plot. The root mean square (RMS) roughness obtained from this AFM image is 7.5 nm. This value is lower than the RMS reported by Takeuchi *et al.* [14], additionally this AFM scan is done on an area that is 25 times larger than the area that is previously reported [14]. Also, our film thickness (500 nm) is orders of magnitude thicker than the 25 nm one reported in [14].

Upon repeating the AFM scans, the same large area (5 X 5 µm$^2$), on the rough films (SEM images in Fig. 2(a) and Fig. 4(a)) gave us the plots that are shown in Fig. 5(b). From the figure, large and very broad peaks and deep valleys affirm the reason behind the film being cloudy and coarser. The RMS value for this case is 148 nm; this value is orders of magnitude higher than the films that are deposited in the method that we proposed. Also, we found that these films will become somewhat rougher as the thickness is increased from 500 nm to 750 nm.

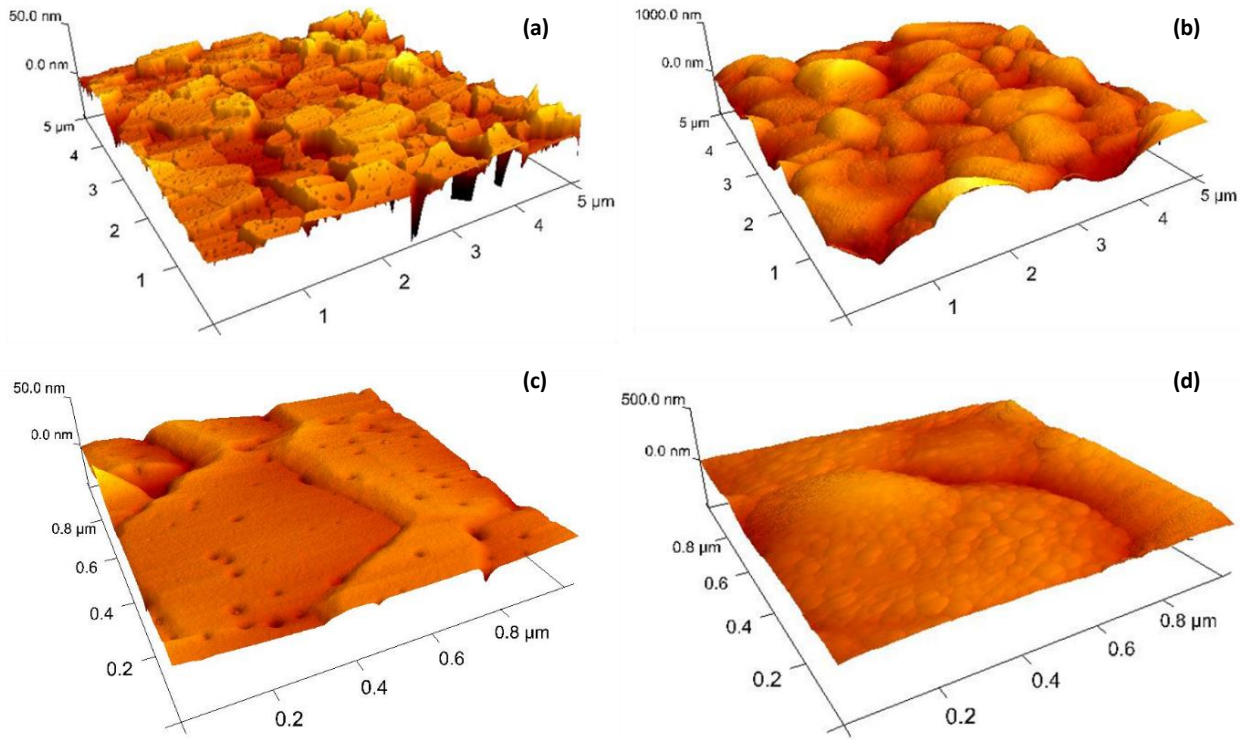

**Figure 5. (a) (5 X 5 µm² area) and (c) (1 X 1 µm² area) are 3D AFM scans of the surface of evaporated smooth films, deposited on glass substrates at optimized rates of 2.6-2.8 nm/s. (b) (1 X 5 µm² area) and (d) (1 X 1 µm² area) are 3D AFM scans of the surface of evaporated rough films, deposited on glass substrates at 0.05-0.08 nm/s.**

A relatively smaller area (1 X 1 µm$^2$) AFM scans were also performed to get a good quantitative estimation of the surface roughness on both optically smooth and rough Sn films. The AFM scan of a smooth film is shown in Fig. 5(c). The RMS value for area scan dropped to even lower values of 3.12 nm, this low RMS value for Sn thin film is not reported anywhere yet.

**Fabrication of optically smooth Sn thin films**

A very similar relatively smaller area AFM scan on the rough Sn film is shown in Fig. 5(d). Although the surface appears to be slight smoother when compared to Fig. 5(b), the RMS value for this scan is 30.8 nm which is still an order magnitude higher than the one shown in Fig. 5(c).

A plot of the RMS roughness as a function of the deposition rate is shown in Fig. 6 and is based on AFM data taken from all four samples, shown in Fig.2. For deposition rate values higher that the highest shown in the figure, the roughness increases dramatically.

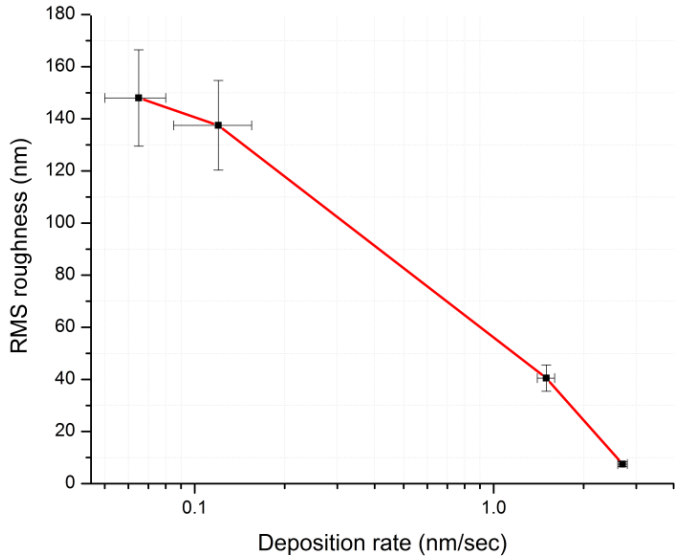

**Figure 6. RMS roughness vs. deposition rate, shown in semi-log scale. The roughness data are obtained from multiple 5 X 5 µm2 AFM scans for each deposition rate case.**

We performed an EDS analysis to confirm the purity of the Sn in the deposited smooth films. Within the sensitivity of the method (a fraction of a percent or so) no other elements are found as can be seen from Fig. 7.

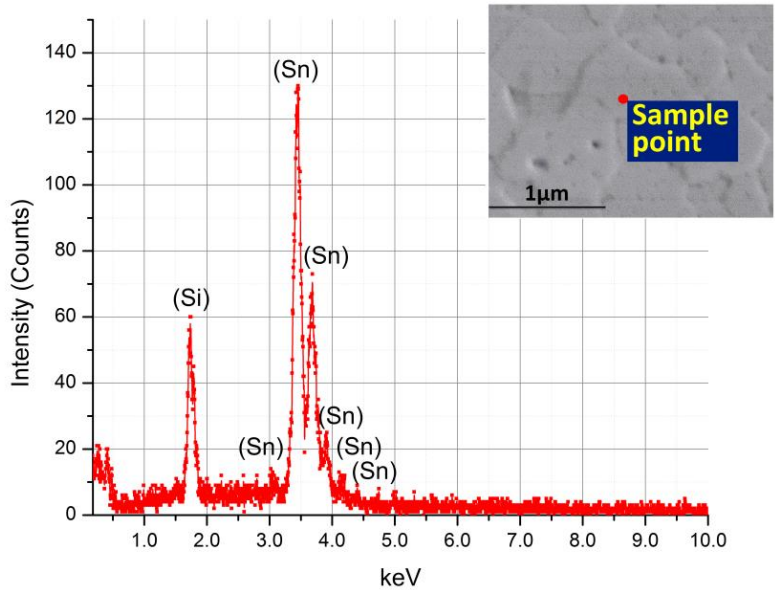

**Figure 7. EDS Spectrum of smooth Sn film (deposited at optimized rates of 2.6-2.8 nm/s) showing the presence of Sn on Si substrate.**

Clearly, we were able to obtain optically smooth Sn films by relatively simple optimization of the deposition rate and the substrate temperature in a basic evaporation set-up. The films were found to be smooth enough for successful excitation of SPPs by using the Otto-prism method[1][2][3]. This relatively simple approach for obtaining smooth Sn films is likely to be of interest to other applications and groups as well [13].

## 4. Conclusions

A method to obtain optically smooth Sn films by vacuum evaporation is established by optimizing the deposition rate and by maintaining the substrate temperature at the room temperature level or below. The fabricated Sn films were characterized by XRD, SEM, AFM and EDS. The AFM work on the Sn films obtained under optimized conditions showed orders of magnitude better smoothness than films obtained under general conditions. The XRD measurements showed that the deposited films have a relatively high preferred orientation with the a-axis of the β-Sn phase being perpendicular to the substrate surface.

**Acknowledgments**

This work was supported by the University of Toledo's EECS department. We also acknowledge helpful discussions with Dr. Eric Tavenner (Creative Polymers; Queensland, Australia) and Dr. Patrick Lavery (Vicor Corporation; North Andover, Massachusetts) and Dr. Victor Karpov

# Fabrication of optically smooth Sn thin films

**List of Figures:**